\documentclass[12pt,a4paper]{article}

\usepackage{amsmath}
\usepackage{amssymb}
\usepackage{epsfig}
\usepackage{graphicx}
\usepackage{cite}

\newcommand{\capdef}{}
\newcommand{\mycaption}[2][\capdef]{\renewcommand{\capdef}{#2}%
       \caption[#1]{{\footnotesize #2}}}
\makeatletter
\renewcommand{\fnum@table}{\textbf{\tablename~\thetable}}
\renewcommand{\fnum@figure}{\textbf{\figurename~\thefigure}}

\renewcommand{\section}{\@startsection{section}{1}{0em}{-\baselineskip}%
{\baselineskip}{\normalfont\large\bfseries}}
\renewcommand{\subsection}%
{\@startsection{subsection}{2}{0em}{-0.7\baselineskip}%
{0.7\baselineskip}{\normalfont\bfseries}}

\newcommand{\stheta}{\sin^22\theta_{13}}

\newcommand{\deltacp}{\delta_\mathrm{CP}}

\begin{document}

\renewcommand{\thefootnote}{\alph{footnote}}

\begin{flushright}
CERN-PH-TH/2006-266\\
\end{flushright}

\vspace*{1.5cm}

\renewcommand{\thefootnote}{\fnsymbol{footnote}}
\setcounter{footnote}{-1}

{\begin{center}
{\large\textbf{%
What is the probability that $\theta_{13}$ and CP violation will be
discovered in future neutrino oscillation experiments?}}
\end{center}}

\renewcommand{\thefootnote}{\it\alph{footnote}}

\vspace*{.8cm}

{\begin{center}{{\bf Thomas Schwetz} $^*$\footnote{$^*$ email: 
\tt schwetz AT cern.ch}}
 \end{center}}
{\it
 \begin{center}
    CERN, Physics Department, Theory Division, CH-1211 Geneva 23, Switzerland
\end{center}}

\vspace*{0.8cm}

\begin{abstract}
  The sensitivity of future neutrino oscillation experiments is
  determined within a frequentist framework by using a statistical
  procedure based on Monte Carlo simulations. I consider the search
  for a non-zero value of the mixing angle $\theta_{13}$ at the T2K
  and Double-Chooz experiments, as well as the discovery of CP
  violation at the example of the T2HK experiment. The probability
  that a discovery will be made at a given confidence level is
  calculated as a function of the true parameter values by generating
  large ensembles of artificial experiments. The interpretation of the
  commonly used sensitivity limits is clarified.
\end{abstract}

\newpage

\renewcommand{\thefootnote}{\arabic{footnote}}
\setcounter{footnote}{0}

\section{Introduction}

The investigation and comparison of the physics potential of future
neutrino oscillation facilities has by now become an industry.
Extensive studies of sensitivities of future experiments to neutrino
parameters are performed, see for example~\cite{ISS}. A widely used
procedure for such calculations---in the following referred to as {\it
standard procedure}---is to assume some input values (``true values'')
for the oscillation parameters for which the predictions for the
observables in a given experiment are calculated without statistical
fluctuations. Then these predictions are used as ``data'' and a
statistical analysis of these data is performed to see how well the
input values for the parameters can be reconstructed by the
experiment. This procedure should give the sensitivity of an
``average'' experiment, where ``average'' lacks a precise definition.

In this letter I will clarify the correct interpretation of such
sensitivities. Focusing on the sensitivity to a non-zero value of the
lepton mixing angle $\theta_{13}$ the potential of a future experiment
will be quantified by answering the following question: 
\begin{quote}
\it Given a true value of $\theta_{13}$, what is the probability that
the hypothesis $\theta_{13} = 0$ can be excluded at a certain
confidence level?
\end{quote} 
This generalises the usual sensitivity limits to a well defined
statistical statement and will allow also a precise definition of the
term ``average experiment''. For example one may define the
sensitivity of an average experiment as the value of
$\theta_{13}^\mathrm{true}$ for which $\theta_{13}=0$ can be excluded
with a probability of 50\%. 

As realistic examples I will consider the T2K~\cite{t2k} and
Double-Chooz~\cite{DC} (D-Chooz) experiments. Details of the
simulation and assumptions on the experimental configurations are
given in Sec.~\ref{sec:exps}. From the statistical point of view there
is an important difference between the two settings. In T2K one looks
for the appearance of $\nu_e$ events from the $\nu_\mu\to\nu_e$
oscillation channel on top of the intrinsic $\nu_e$ background in the
beam. Therefore it is similar to the problem of a Poisson signal over
a background. In contrast, D-Chooz is a disappearance experiment at a
nuclear reactor comparing the $\bar\nu_e$ signal in a near and far
detector where in both detectors a very large number of events is
obtained, and one looks for a small difference in these large numbers
beyond the geometrical scaling with the distance. Hence one may expect
that in this case Gaussian approximations are well justified.

In Sec.~\ref{sec:th13} the sensitivity of T2K and D-Chooz to $\stheta$
is considered by performing a Monte Carlo (MC) simulation of the
experiments. A large number of artificial data sets is generated to
calculate the actual distribution of the statistics used to decide
whether $\theta_{13}=0$ should be rejected at a given confidence level
(CL). This will allow to answer the question stated above within a
well defined frequentist framework.
Moreover one does not rely on questionable assumptions necessary in
the standard procedure, for example issues related to the non-linear
character of the parameters, the periodicity of the CP phase
$\deltacp$, the physical boundary $\stheta \ge 0$, assuming standard
$\chi^2$-distributions, and the question of how many degrees of
freedom to use for them.

In Sec.~\ref{sec:CP} I extend these methods to the case of CP
violation searches, where the example of the T2HK experiment is
considered. This case is especially interesting from the statistical
point of view, since the quantity of interest ``CP violation'' has
rather non-standard statistical properties since it is directly
related to the periodic variable $\deltacp$, and the assumption of a
linear parameter dependence of the observables which is implied by the
use of standard $\chi^2$-distributions is not justified a priori.

\section{Description of the experiment simulations}
\label{sec:exps}

For the simulation of the D-Chooz reactor experiment the information
available in the proposal Ref.~\cite{DC} is used. For the far detector
at distances of 998~m and 1115~m from the two reactor cores an
exposure of 5 years with 60.5\% efficiency is assumed. The near
detector at an equal distance of 280~m from both cores will come
online somewhat later and I take 3 years of data with 43.7\%
efficiency. This gives in total $75\,000$ events in the far and
$473\,400$ events in the near detector~\cite{DC}. I take into account
an uncertainty on the reactor neutrino flux of 2\% (uncorrelated
between the two cores) and assume an error in the relative
normalisation of the two detectors of 0.6\%. Furthermore I include a
background of 3.6\% (2.7\%) in the far (near) detector which is known
within an uncertainty of 20\%. A fit is performed for the two
oscillation parameters $\stheta$ and $\Delta m^2_{31}$, where always
the true value $\Delta m^2_{31} = 2.5\times 10^{-3}$~eV$^2$ is adopted
and external information on $\Delta m^2_{31}$ with an accuracy of 5\%
at 1$\sigma$ is assumed. More details of the reactor simulation can be
found in Ref.~\cite{Huber:2003pm} and the standard limits are in good
agreement with Ref.~\cite{DC}.

For the generation of artificial data for the D-Chooz experiment
I assume that the systematical uncertainties are random variables,
i.e., for the flux normalisations from the two reactor cores and for
the normalisations of the two detectors Gaussian variables are thrown
with the errors given above. Then the expected event number in each
bin of each detector is shifted accordingly, and this shifted value is
used as mean for generating the ``observed'' event number in this bin
according to a Poisson distribution.

For the simulation of the T2K and T2HK experiments I follow closely
the setup provided within the GLoBES software package~\cite{globes}
which is based on information from Ref.~\cite{t2k}. However, in order
to be able to perform the MC simulation for the analysis presented
here a dedicated code has been developed which drastically reduces the
required calculation time. To this aim the following simplifications
have been adopted. The oscillation parameters $\sin^2\theta_{12}$,
$\Delta m^2_{21}$, and $\Delta m^2_{31}$ are fixed to their assumed
true values $0.3$, $7.9\times 10^{-5}$~eV$^2$, and $2.5\times
10^{-3}$~eV$^2$, respectively. I analyse only the appearance channel,
the disappearance channel is taken into account implicitly by
fixing $\Delta m^2_{31}$ and assuming an external uncertainty on
$\sin^2\theta_{23}$ of $0.08$ ($0.04$) at $1\sigma$ for T2K
(T2HK). The true value of $\sin^2\theta_{23}$ is assumed to be $0.5$,
which implies that the octant degeneracy is absent. I do not take into
account the sgn($\Delta m^2_{31}$) degeneracy and assume always normal
neutrino mass hierarchy.

For the T2K simulation I assume 5 years of data taking in the neutrino
channel, a 0.76~MW beam, and the 22.5~kt fiducial mass of the SK
detector. Signal and background events are normalised to the standard
GLoBES setup~\cite{globes} and systematical errors of 5\% are assumed
on signal and background. Despite the simplifications mentioned above
standard sensitivity limits of this T2K simulation are in excellent
agreement with the ones calculated with GLoBES. 
The T2HK implementation follows the setup adopted in
Ref.~\cite{Campagne:2006yx}, which consists of a 4~MW beam, the HK
detector with 440~kt mass, 2 years of neutrino and 8 years of
anti-neutrino data. Uncorrelated systematical errors of 5\% are
included for the signal and background in neutrino and anti-neutrino
running. In the case of T2HK minor differences of standard sensitivity
limits appear between the simulation used here and the GLoBES
implementation due to the adopted simplifications and other subtle
differences in the analysis.

Artificial data sets for T2K and T2HK are generated in the following
way. Assuming true values of $\theta_{13}$ and $\deltacp$ the
predicted spectrum in reconstructed neutrino energy is calculated
using a Gaussian energy resolution of 85~MeV due to Fermi motion for
quasi-elastic (QE) events and no energy information for non-QE events
(i.e., energy smearing with uniform distribution). The QE and non-QE
event spectra are added taking into account the relevant ratio of
cross sections. Systematical uncertainties are considered as random
variables, i.e., for each systematic a Gaussian variable is thrown
with an error of 5\% and the signal and background spectra are
rescaled accordingly. Then the ``observed'' number of events in each
bin is generated from a Poisson distribution with the mean
corresponding to the systematic-shifted predicted event numbers.

\section{Sensitivity to $\theta_{13}$}
\label{sec:th13}

In this section I am going to answer the question quoted in the
introduction considering the T2K and D-Chooz experiments. I start
by describing in some detail the procedure used for this purpose.

\begin{figure}[!t]
\centering 
\includegraphics[width=0.6\textwidth]{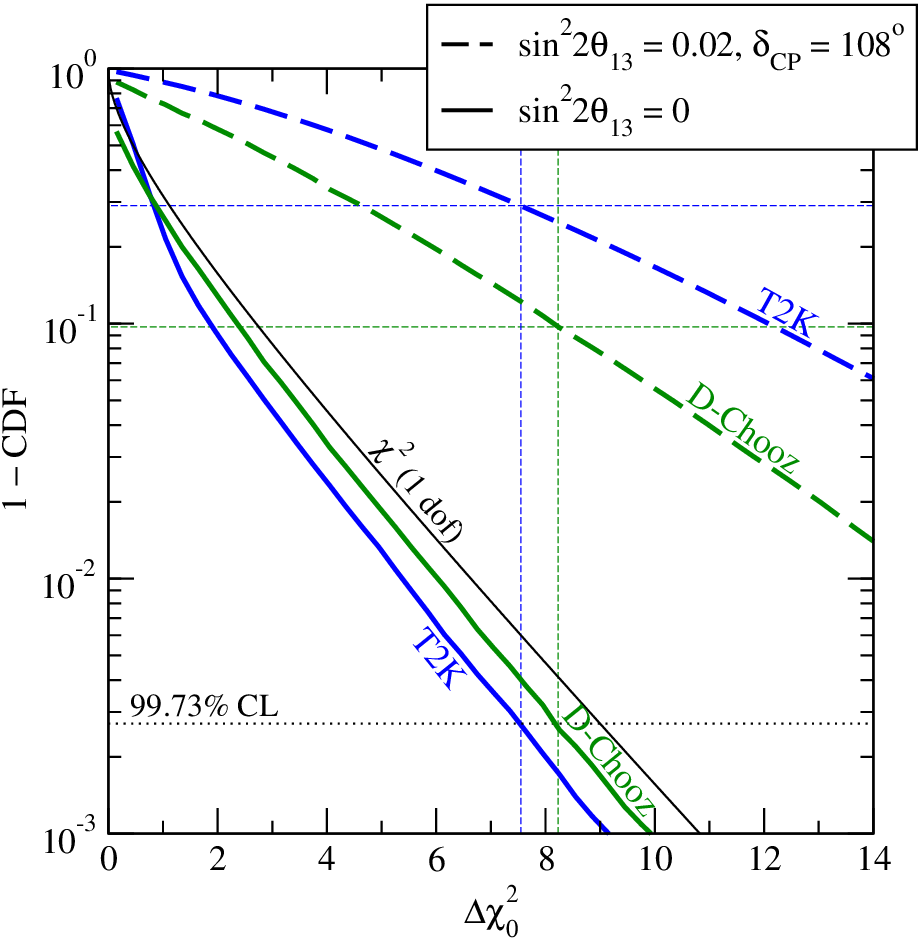}
  \mycaption{Illustration how the probability is obtained that T2K and
  D-Chooz will discover the value $\stheta = 0.02$ at 99.73\%~CL. The
  solid (dashed) curves correspond to the CDF of the distribution
  $f_0$ ($f_{\theta_{13}}$) of $\Delta\chi^2_0$ generated for a value
  $\theta_{13} = 0$ ($\stheta = 0.02$ and $\deltacp = 108^\circ$). For
  comparison also the CDF of a $\chi^2$-distribution for 1~degree of
  freedom is shown. The vertical and horizontal lines indicate how the
  probability $P_\mathrm{disc}$ is obtained, see text for details.}
\label{fig:dist}
\end{figure}

First one has to define a criterion to decide whether (real or
artificial) data are consistent with the hypothesis $\theta_{13} =
0$. I will use a test based on the likelihood ratio
\begin{equation}\label{eq:LHratio}
2 \ln \frac{\mathcal{L}_\mathrm{max}}{\mathcal{L}(\theta_{13} = 0)}
=
\chi^2(\theta_{13} = 0) - \chi^2_\mathrm{min} 
\equiv \Delta\chi^2_0 \,,
\end{equation}
where the relation $\chi^2 = -2 \ln\mathcal{L}$ between the $\chi^2$
and the likelihood function of the data has been used.
$\chi^2_\mathrm{min} = -2 \ln\mathcal{L}_\mathrm{max}$ are the
corresponding values at the best fit point which are compared to the
values at $\theta_{13} = 0$. If the statistic (\ref{eq:LHratio}) is
larger than a value $\lambda(\alpha)$ the hypothesis $\theta_{13}=0$
can be excluded at the $100(1-\alpha)$\%~CL. The value
$\lambda(\alpha)$ is calculated by MC in the following way. Assuming
$\theta_{13} = 0$ many artificial data sets are generated. In other
words, an ensemble of many repeated experiments is considered assuming
that the true value of $\theta_{13}$ is zero. For each artificial data
set the $\chi^2$ is minimised to obtain the distribution
$f_0(\Delta\chi^2_0)$ of the statistic (\ref{eq:LHratio}).  Given this
distribution $\lambda(\alpha)$ is defined by the requirement that a
fraction $\alpha$ of all experiments will have $\Delta\chi^2_0 >
\lambda(\alpha)$:
\begin{equation}
\alpha = \int_{\lambda(\alpha)}^\infty dx \, f_0(x) \,.
\end{equation}
The cumulative distribution function (CDF) of $f_0$ is shown in
Fig.~\ref{fig:dist} as solid curves for T2K and D-Chooz. In the
Gaussian approximation $\Delta\chi^2_0$ should be distributed
according to a $\chi^2$-distribution for 1~degree of freedom. As
visible in Fig.~\ref{fig:dist} for the examples under consideration
there are some deviations from this situation, where the difference is
larger for T2K. The cut value $\lambda(\alpha)$ for the 99.73\%~CL
(which is equal to 9 for the $\chi^2$-distribution) is 8.23 for
D-Chooz and 7.55 for T2K.

If the experiment had been performed already and real data were
available one would now check if $\chi^2(\theta_\mathrm{13} = 0) -
\chi^2_\mathrm{min,data}$ is larger than $\lambda(\alpha)$ to decide
whether the hypothesis $\theta_{13} = 0$ can be
excluded.\footnote{Note that the test for the hypothesis $\theta_{13}
= 0$ used here is equivalent to ask whether the point $\theta_{13} =
0$ is contained in the $100(1-\alpha)\%$~CL region constructed
according to the Feldman--Cousins prescription~\cite{Feldman:1997qc}.}
In the absence of real data one can, however, calculate the
probability for this to occur as a function of the value of
$\theta_{13}$. More precisely, assuming a fixed value of $\theta_{13}$
(and in case of T2K also of $\deltacp$) many artificial data sets are
generated. This yields the distribution
$f_{\theta_{13}}(\Delta\chi^2_0)$ under the assumption of the ``true
value'' $\theta_{13}$. The probability $P_\mathrm{disc}(\alpha,
\theta_{13})$ that $\theta_{13}=0$ can be excluded at the
$100(1-\alpha)\%$~CL is given by
\begin{equation}\label{eq:PdiscFC}
P_\mathrm{disc} (\alpha, \theta_{13})\equiv 
P\left[\Delta\chi^2_0 > \lambda(\alpha) \: | \: \theta_{13}\right]
= \int_{\lambda(\alpha)}^\infty dx\, f_{\theta_{13}}(x) \,.
\end{equation}
The calculation of $P_\mathrm{disc}$ is illustrated in
Fig.~\ref{fig:dist} assuming that the true value of $\stheta$ is
0.02. One can read off from this figure that the probability to
exclude the hypothesis $\theta_{13} = 0$ at 99.73\%~CL is about 29\%
for T2K (if $\deltacp = 108^\circ$) and 9.7\% for D-Chooz.

\begin{figure}[!t]
\centering 
\includegraphics[width=0.7\textwidth]{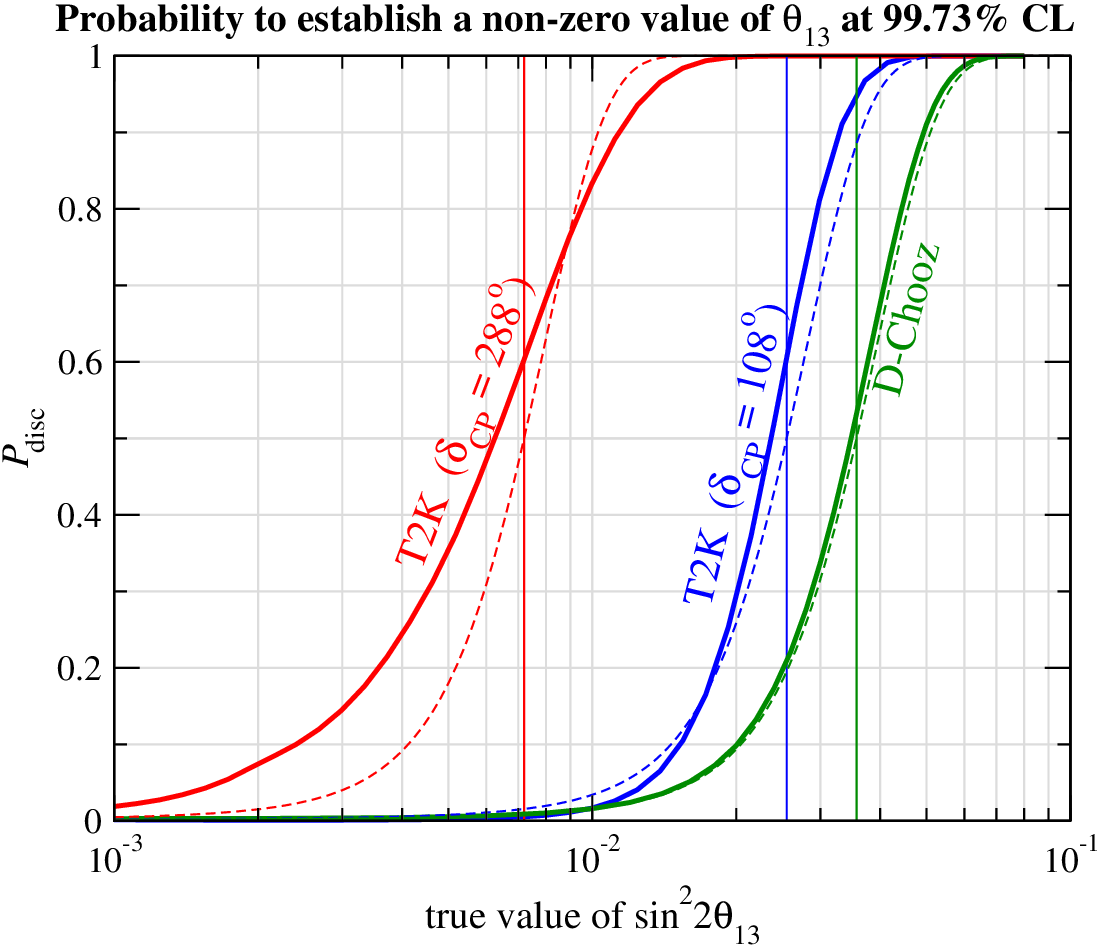}
  \mycaption{The probability to exclude the hypothesis $\theta_{13} =
  0$ at the 99.73\%~CL for T2K and D-Chooz as a function of the true
  value of $\stheta$. The two curves for T2K correspond to the true
  values $\deltacp = 108^\circ$ and $288^\circ$. The vertical lines
  show the corresponding ``standard'' sensitivities. The dashed curves
  correspond to the probability $P_\mathrm{disc}$ calculated in the
  Gaussian approximation according to Eq.~(\ref{eq:gauss}).}
\label{fig:th13-prob}
\end{figure}

Now one has to scan over the true values of $\theta_{13}$ (and in the
case of T2K also over $\deltacp$), repeating the procedure outlined
above in each point.  Fig.~\ref{fig:th13-prob} shows the probability
$P_\mathrm{disc}$ to exclude the hypothesis $\theta_{13} = 0$ at the
99.73\%~CL for T2K and D-Chooz as a function of the true value of
$\stheta$. For each true value $3\times 10^6$ data sets have been
simulated. For T2K the two values chosen for $\deltacp$ correspond
roughly to the best and worst sensitivity. The vertical lines in the
plot show the standard sensitivities calculated from the condition
$\Delta \chi^2 \ge 9$ without statistical fluctuations. One observes
that for D-Chooz the standard sensitivity corresponds indeed with good
accuracy to $P_\mathrm{disc} = 50\%$, as expected for an ``average''
experiment. For T2K the discovery probabilities corresponding to the
standard sensitivities are actually slightly higher, around 60\%.

The dashed curves shown in Fig.~\ref{fig:th13-prob} are obtained
assuming a Gaussian measurement of $\stheta$. In this case
$P_\mathrm{disc}$ can be obtained in terms of the error function in the
following way. Assuming that $x$ is a Gaussian variable with standard
deviation $\sigma$ the hypothesis $x = 0$ can be excluded at the
99.73\%~CL if the observed value $x^\mathrm{obs}$ is bigger than
$3\sigma$. On the other hand, the probability for $x^\mathrm{obs} \ge
3\sigma$ as a function of the true value $x^\mathrm{true}$ is easily
calculated as
\begin{equation}\label{eq:gauss}
P\left[ x^\mathrm{obs} \ge 3\sigma \:|\: x^\mathrm{true} \right] =
\int_{3\sigma}^\infty dx \, G(x; \: x^\mathrm{true}, \sigma) =
\frac{1}{2} \left[ 1 - \mathrm{erf}
\left(\frac{3\sigma - x^\mathrm{true}}{\sqrt{2}\sigma}\right)\right] \,,
\end{equation}
where $G(x; \: x^\mathrm{true}, \sigma)$ denotes the normal
distribution with mean $x^\mathrm{true}$ and standard deviation
$\sigma$. 

The dashed curves in Fig.~\ref{fig:th13-prob} have been obtained from
Eq.~(\ref{eq:gauss}) by identifying $\stheta = x$ and by using for
$\sigma$ one third of the 99.73\%~CL sensitivity limit from the
standard procedure. One observes that for D-Chooz this approximation
is excellent. Hence, in this case $\stheta$ can be considered indeed
as a Gaussian variable and the probability $P_\mathrm{disc}$ can be
calculated from the standard sensitivity limit and
Eq.~(\ref{eq:gauss}) without the need of a MC simulation.
In contrast, for T2K some deviations from Gaussianity are visible
(especially for $\deltacp^\mathrm{true} = 288^\circ$). This is not
unexpected, since in this case event numbers are small, background
fluctuations are important, and the dependence of the observables on
the parameters is much more complicated than in the case of D-Chooz.

\begin{figure}[!t]
\centering 
\includegraphics[width=0.8\textwidth]{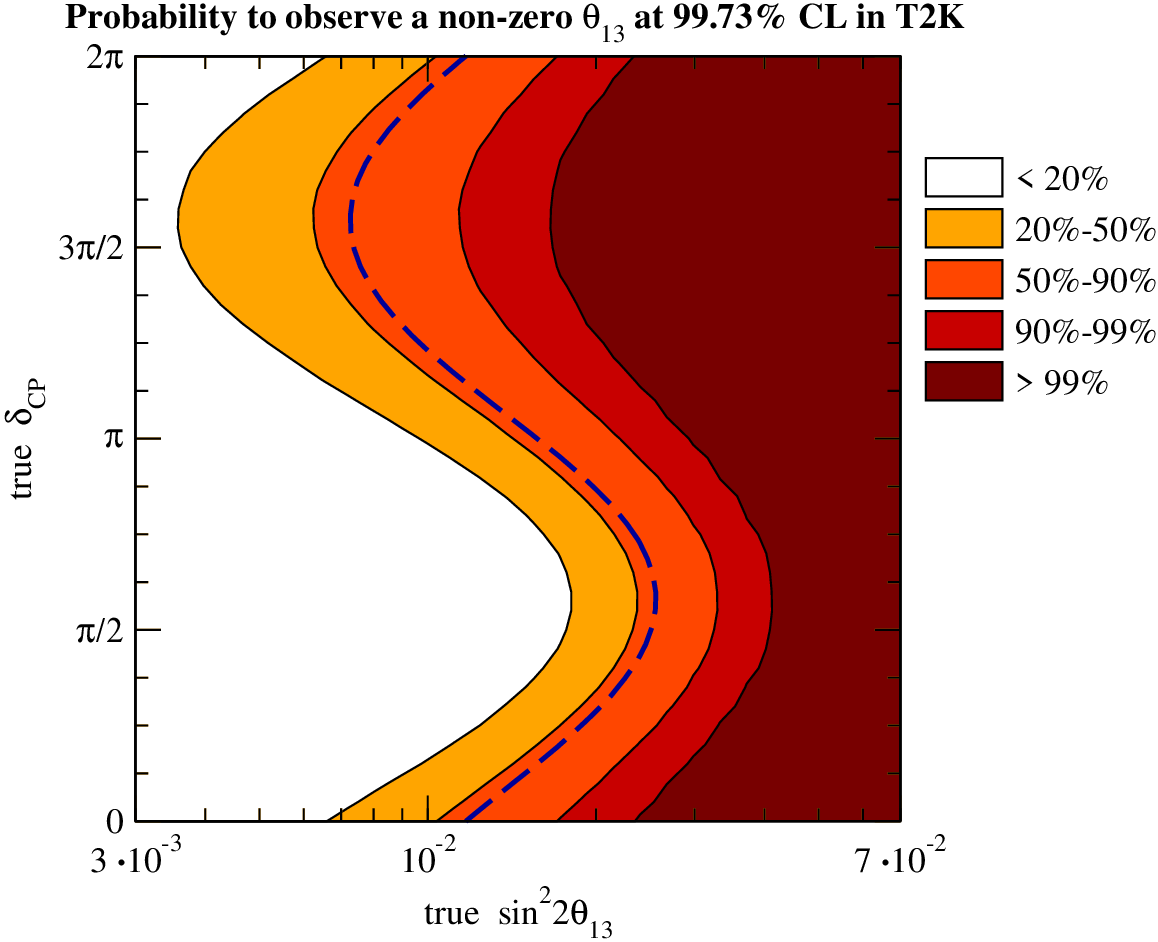}
  \mycaption{Contours of the probability $P_\mathrm{disc}$ to
  establish a non-zero value of $\theta_{31}$ at the 99.73\%~CL for
  T2K in the $\stheta^\mathrm{true}$-$\deltacp^\mathrm{true}$ plane.
  The dashed curve corresponds to the ``standard sensitivity limit''.}
\label{fig:th13-contours}
\end{figure}

Contours of the probability $P_\mathrm{disc}$ for the T2K experiment
in the plane of $\stheta^\mathrm{true}$ and $\deltacp^\mathrm{true}$
are shown in Fig.~\ref{fig:th13-contours}. $P_\mathrm{disc}$ has been
calculated for a grid of $41\times 41$ values and at each point in the
grid $10^5$ data sets have been generated, leading in total to nearly
$1.7\times 10^8$ performed fits. This figure is the generalisation of
the usual sensitivity limit (shown as dashed curve) and for each true
value of the parameters one can infer the probability that T2K can
establish a non-zero value of $\theta_{13}$ at the 99.73\%~CL. As
indicated already in Fig.~\ref{fig:th13-prob} one finds that the
standard sensitivity curve corresponds roughly to a discovery
probability of 60\%. The region where $\theta_{13} > 0$ can be
established with high probability, let's say greater than 99\%, is
found for $\stheta^\mathrm{true} > 0.0166 - 0.041$, depending on the
true value of $\deltacp$. It is shifted with respect to the standard
sensitivity limit to values of $\stheta$ larger by roughly a factor of
2.

\section{Sensitivity to CP violation}
\label{sec:CP}

In this section I will apply similar concepts to the sensitivity to CP
violation (CPV), using as example the T2HK experiment. Now the
relevant question is:
\begin{quote}
\it Given true values of $\theta_{13}$ and $\deltacp$, what is the
probability that CPV can be established at a certain confidence level?
\end{quote} 
In this case the hypothesis to be excluded, CP conservation (CPC), is
more complicated than in the case of the $\theta_{13}$
sensitivity. Whereas in the previous case the hypothesis was just a
single point in the parameter space ($\theta_{13} = 0$, independent of
$\deltacp$), now one wants to exclude $\deltacp = 0$ and $\deltacp =
\pi$ for any value of $\theta_{13}$. Therefore, in the following the
phrase ``establish CPV at the $100(1-\alpha)\%$~CL'' is considered to
be equivalent to the statement that for any value of $\stheta$
neither $\deltacp = 0$ nor $\deltacp = \pi$ is contained in the
confidence regions in the $\stheta$-$\deltacp$ plane at the
$100(1-\alpha)\%$~CL. For the construction of the confidence regions
the Feldman--Cousins method~\cite{Feldman:1997qc} will be used .

\begin{figure}[!t]
\centering 
\includegraphics[width=0.6\textwidth]{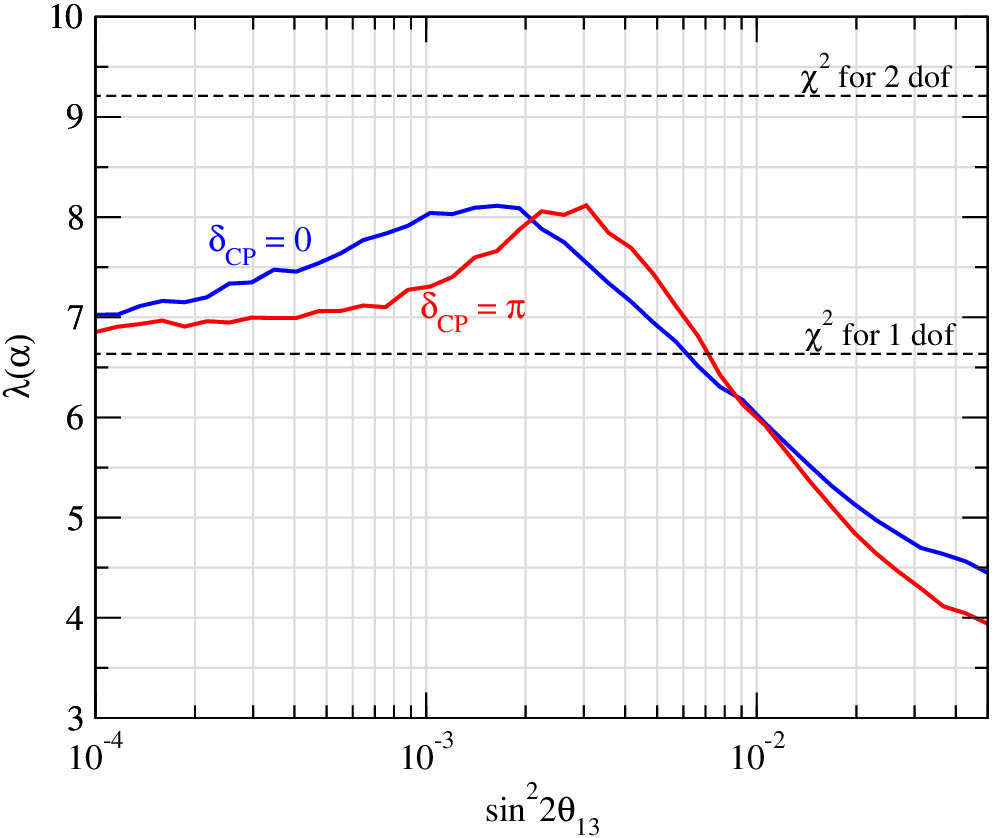}
  \mycaption{The cut value $\lambda(\alpha; \,
  \theta_{13},\delta_\mathrm{CPC})$, see Eqs.~(\ref{eq:CPVtest}) and
  (\ref{eq:alphaCP}), for the 99\%~CL ($\alpha = 0.01$) as a function
  of $\stheta$ for $\deltacp = 0$ and $\pi$. The horizontal lines show
  the canonical cut values following from $\chi^2$-distributions for 1
  and 2~degrees of freedom.}
\label{fig:Dchisq}
\end{figure}

This method is implemented as follows. To decide whether given data
are consistent with CPC at the $100(1-\alpha)\%$~CL it is checked if
the statistic
\begin{equation}
\Delta\chi^2_\mathrm{CPC}(\theta_{13}, \delta_\mathrm{CPC}) 
\equiv \chi^2(\theta_{13}, \delta_\mathrm{CPC}) -
\chi^2_\mathrm{min} 
\quad\text{with}\quad 
\delta_\mathrm{CPC} = 0, \pi
\end{equation}
fulfils the condition
\begin{equation}\label{eq:CPVtest}
\Delta\chi^2_\mathrm{CPC}(\theta_{13}, \delta_\mathrm{CPC}) > 
\lambda(\alpha; \, \theta_{13},\delta_\mathrm{CPC})
\end{equation}
for all values of $\theta_{13}$, where $\lambda(\alpha;\,
\theta_{13},\delta_\mathrm{CPC})$ is calculated by MC simulation. Many
artificial data sets are generated for the parameters $\deltacp =
0,\pi$ and $\theta_{13}$, to map out the distribution
$f_\mathrm{CPC}(\Delta\chi^2_\mathrm{CPC} \:|\: \theta_{13},
\delta_\mathrm{CPC})$. Then $\lambda(\alpha; \,
\theta_{13},\delta_\mathrm{CPC})$ is determined by\footnote{Note that
in the case of the $\theta_{13}$ sensitivity the corresponding
distribution $f_0$, and therefore also $\lambda(\alpha)$ are
independent of the true parameter values. In the case of the CPV
sensitivity the explicit parameter dependence makes it necessary to
refer to the full Feldman--Cousins confidence region construction,
whereas for the $\theta_{13}$ sensitivity this procedure is equivalent
to a simple likelihood ratio test.}
\begin{equation}\label{eq:alphaCP}
\alpha = \int_{\lambda(\alpha; \, \theta_{13},\delta_\mathrm{CPC})}^\infty 
dx \, f_\mathrm{CPC}(x \:|\: \theta_{13}, \delta_\mathrm{CPC})\,.
\end{equation}
In Fig.~\ref{fig:Dchisq} $\lambda(\alpha; \,
\theta_{13},\delta_\mathrm{CPC})$ is shown for $\alpha = 0.01$.  In
contrast to the values 6.635~(9.21) following from
$\chi^2$-distributions for 1~(2)~degrees of freedom the actual cut
values vary between 4 and 8.1 in the considered range of $\stheta$.

\begin{figure}[!t]
\centering 
\includegraphics[width=0.8\textwidth]{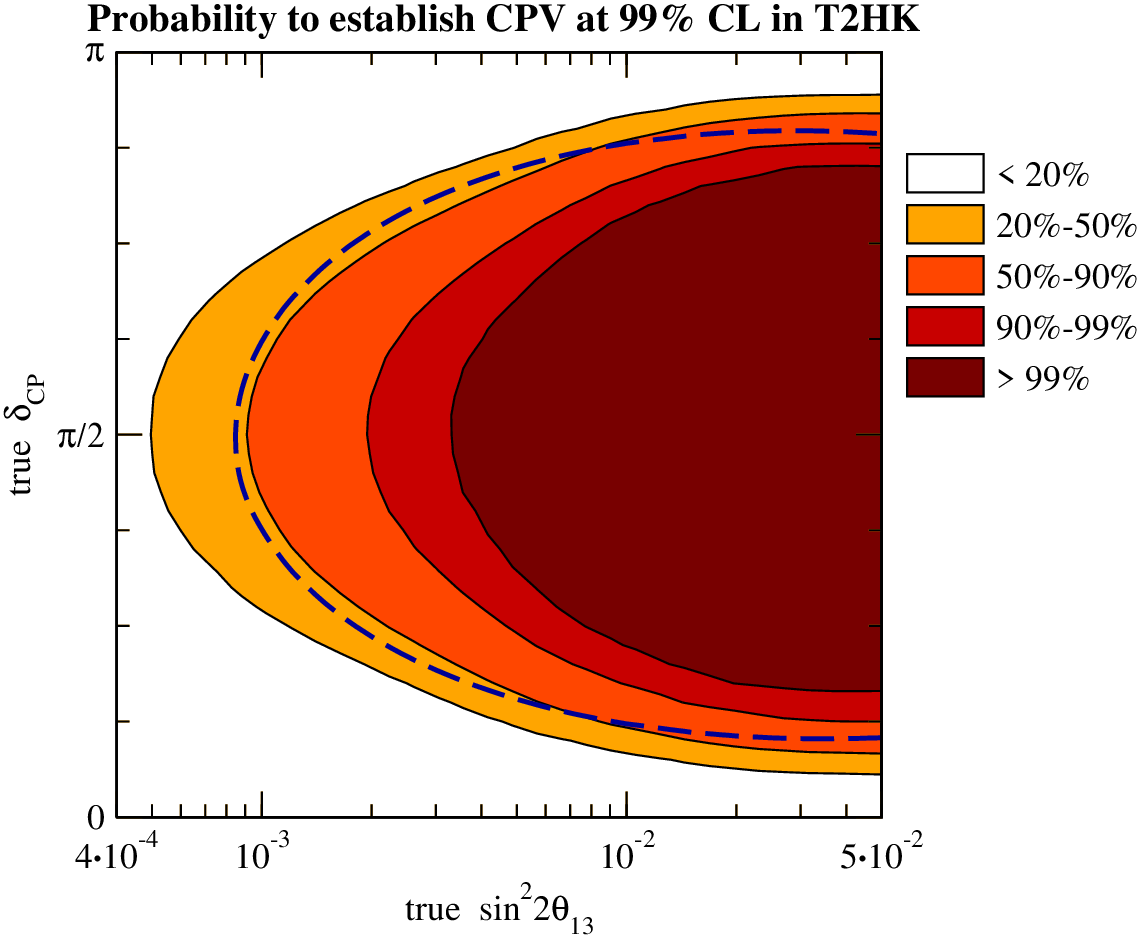}
  \mycaption{Contours of the probability $P_\mathrm{disc}$ to
  establish CPV at the 99\%~CL for T2HK in the
  $\stheta^\mathrm{true}$-$\deltacp^\mathrm{true}$ plane. The dashed
  curve corresponds to the ``standard sensitivity limit''.}
\label{fig:CP}
\end{figure}

To determine the probability $P_\mathrm{disc}$ that CPV can be
established for given true values of $\theta_{13}$ and $\deltacp$ many
artificial data sets are generated under the assumption of these
parameters. For each data set the test (\ref{eq:CPVtest}) is performed
and $P_\mathrm{disc}$ is given by the fraction of data sets for which
CPV can be established at the corresponding CL. 
The results of such an analysis for the T2HK experiment are shown in
Fig.~\ref{fig:CP}. $P_\mathrm{disc}$ has been calculated for a grid of
$41\times 41$ true values and at each point in the grid $5\times 10^4$
data sets have been generated. This number is somewhat reduced with
respect to the one used for the $\theta_{13}$ sensitivity to keep the
analysis feasible. Because of the smaller ensemble the CL is reduced
to 99\%. The plot is restricted to the interval $0\le \deltacp \le
\pi$, similar behaviour is expected for $\pi < \deltacp < 2\pi$.

Fig.~\ref{fig:CP} provides the complete information on the
possibilities to discover CPV in T2HK. For each true value of the
parameters one can infer the probability that CPV can be establish at
the 99\%~CL. The standard sensitivity limit (obtained from the 
condition $\Delta \chi^2 \ge 6.635$ without taking into account
statistical fluctuations) is shown as dashed curve. It is remarkable
that the standard limit is rather close to the contour for
$P_\mathrm{disc} = 50\%$, i.e., an ``average'' experiment. The
deviations from this value can be motivated from
Fig.~\ref{fig:Dchisq}. Since $\lambda(\alpha; \,
\theta_{13},\delta_\mathrm{CPC})$ is bigger than the canonical value
6.635 for $\stheta \lesssim 10^{-2}$ the sensitivity to CPV is
actually slightly worse leading to $P_\mathrm{disc} < 50\%$ for the
standard limit, whereas for $\stheta \gtrsim 10^{-2}$ one has 
$\lambda(\alpha; \, \theta_{13},\delta_\mathrm{CPC}) < 6.635$ which
leads to better sensitivities to CPV and hence $P_\mathrm{disc} >
50\%$ at the standard limit.

Despite the rather good approximation of the standard sensitivity
limit for $P_\mathrm{disc}= 50\%$ it is evident from Fig.~\ref{fig:CP}
that one has to be aware of the correct interpretation of such limits.
The region where CPV can be established with high probability is
significantly smaller than the standard sensitivity region. For
example, maximal CPV $\deltacp = \pi/2$ can be established with a
probability of more than 99\% at the 99\%~CL for $\stheta > 3.3\times
10^{-3}$, whereas the corresponding standard sensitivity limit is
$\stheta = 8.5\times 10^{-4}$, nearly a factor 4 smaller.

\section{Summary}

In this work the sensitivity of future neutrino oscillation
experiments has been calculated by using a statistical procedure
within a frequentist framework based on MC simulations.  I have
determined the probability that a discovery will be made at a given CL
as a function of the true parameter values by generating a large
ensemble of artificial experiments. The interpretation of the widely
used ``standard'' sensitivity limits, where statistical fluctuations
are neglected, has been clarified.

As examples I have considered the discovery of a non-zero value for
$\theta_{13}$ at the T2K and Double-Chooz experiments. It has been
found that for Double-Chooz the Gaussian approximation is very well
justified. The usually calculated sensitivity corresponds to the
performance of an average experiment (the discovery will be made with
a probability of 50\%), and the actual discovery probability can be
estimated by a simple formula in terms of the error function. In the
case of T2K some deviations from Gaussianity are found and the
standard sensitivity limits correspond to a discovery probability of
about 60\%.
Similar concepts have been applied to the discovery of CP violation in
neutrino oscillations at the T2HK experiment. A definition of
``establishing CP violation'' based confidence regions
according to the Feldman--Cousins prescription has been used.

For all considered cases I have found that standard sensitivity limits
provide a reasonable approximation for an average experiment,
corresponding to a discovery probability of order 50\%. However, one
has to be aware of the correct interpretation of such limits. In
general the region where a discovery can be made with high probability
is significantly smaller than the one corresponding to the standard
sensitivity limits.

\subsection*{Acknowledgement}

This work has been triggered partially by the talk
Ref.~\cite{ConradNuFact} and I thank J.~Conrad for stimulating
communication on the topic of sensitivity limits.

\end{document}